\documentstyle[12pt,epsfig]{article}
\advance \topmargin by -\headheight
\advance \topmargin by -\headsep

\evensidemargin \oddsidemargin
\begin{document}
%10 de abril, 2001
%               macros formatting and equations
\topmargin -.6in
\def\br{\begin{eqnarray}}
\def\er{\end{eqnarray}}
\def\be{\begin{equation}}
\def\ee{\end{equation}}
\def\({\left(}
\def\){\right)}
\def\a{\alpha}
\def\b{\beta}
\def\d{\delta}
\def\D{\Delta}
\def\g{\gamma}
\def\G{\Gamma}
\def\h{ {1\over 2}  }
\def\hp{ {+{1\over 2}}  }
\def\hm{ {-{1\over 2}}  }
\def\k{\kappa}
\def\l{\lambda}
\def\L{\Lambda}
\def\m{\mu}
\def\n{\nu}
\def\o{\over}
\def\O{\Omega}
\def\p{\phi}
\def\rh{\rho}
\def\s{\sigma}
\def\t{\tau}
\def\th{\theta}
\def\ii {\'\i  }

\begin{center}
{\large {\bf A new exactly solvable Eckart-type potential}}\footnotemark
\end{center}
\normalsize
\vskip 1cm
\begin{center}

{\it  Elso Drigo Filho $^a$ \footnotemark
\footnotetext{elso@df.ibilce.unesp.br}
and Regina Maria  Ricotta $^b$} \footnotemark
\footnotetext{regina@fatecsp.br}\\
$^a$ Instituto de Bioci\^encias, Letras e Ci\^encias Exatas, IBILCE-UNESP\\
Rua Cristov\~ao Colombo, 2265 -  15054-000 S\~ao Jos\'e do Rio Preto - SP\\
$^b$  Faculdade de Tecnologia de S\~ao Paulo, FATEC/SP-CEETPS-UNESP  \\
Pra\c ca  Fernando Prestes, 30 -  01124-060 S\~ao Paulo-SP\\
Brazil\\
\vskip 1cm
\end{center}
{\bf  Abstract}\\
A new exact analytically solvable Eckart-type potential is presented,   a
generalisation of the Hulth\'en potential. The study through Supersymmetric Quantum
Mechanics is presented  together with the  hierarchy of Hamiltonians and the shape invariance
property.\\

\noindent   {\bf I. Introduction}\\

Based on the study of SUSY breaking mechanism of higher dimensional quantum field theories
\cite{Witten},  Supersymmetric Quantum Mechanics (SQM) appeared 20 years ago and has so far
been considered as a new field of research, providing not  only a supersymmetric interpretation 
of the Schr\" {o}dinger equation, but important results to a variety of
non-relativistic quantum mechanical problems, \cite{Cooper1}.  Particular examples to be mentioned
include the better understanding of  the exactly solvable, \cite{Gedenshtein}-\cite{Dutt},   the 
partially solvable, \cite{Drigo1}-\cite{ Drigo3}, the isospectral, \cite{Dunne} and   the
periodic potentials, \cite{Sukhatme}. The association of the variational method with SQM  
formalism has been introduced to obtain the approximate energy spectra of  non-exactly solvable
potentials,  \cite{Gozzi}-\cite{Drigo4}. In  the work of reference \cite{Drigo4} a new
methodology, based on  an {\it ansatz} for the superpotential which is  related to the trial wave
function, has been proposed. Using physical arguments it is possible to make an {\it ansatz} for
the superpotential which satisfies the Riccati equation by an effective potential. The
superalgebra enables us to take this superpotential and evaluate the trial wavefunctions
containing the variational parameter, i.e.,  the parameter that minimises the energy expectation
value. This new scheme has been successfully applied to obtain the spectra of 3-dimensional
atomic systems  well fit by the Hulth\'en, the Morse and the  screened Coulomb potentials,
\cite{Drigo4}-\cite{Drigo7}. 

In particular, when applying the approach to the Hulth\'en potential, \cite{Drigo4}, it was found
that its effective potential was linked to a new exactly solvable potential, that presents,
unlike the Hulth\'en potential in one-dimension the property of shape invariance.  It is a two
parameters potential, apart from the screening parameter $\d$ and it has already appeared in the
literature, \cite{Varshni} in a similar form. Written in terms of hyperbolic functions, this 
potential is known as the exactly solvable Eckart potential.

In this letter a study of such new exactly solvable potential through SQM is presented. We show 
its hierarchy and its shape invariance property. For particular values of its constants  the 
Hulth\'en potential hierarchy is recovered.  This material is preceded by a brief review of
SQM, in order to fix the notation.\\

\noindent {\bf  II. Supersymmetric Quantum Mechanics}\\

In SQM for $N=2$ we have two
nilpotent operators, $Q$ and $Q^+$, satisfying the algebra
\be
\{ Q, Q^+\} = H_{SS} \;,\;\;\; Q^2  = {Q^+}^2 = 0,
\ee
where $H_{SS}$ is the supersymmetric Hamiltonian. This algebra can be
realized as
\be
Q =  \left( \begin{array}{cc} 0  & 0  \\ A^-  & 0
\end{array} \right ) \;,\;\;\;
Q^+ = \left( \begin{array}{cc} 0  & A^+  \\ 0 & 0
\end{array} \right )
\ee
where $A^{\pm}$ are bosonic operators.  With this realization the
supersymmetric Hamiltonian
$H_{SS}$ is given by
\be
H_{SS} = \left( \begin{array}{cc} A^+A^-  &  0 \\ 0 & A^-A^+
\end{array} \right ) = \left( \begin{array}{cc} H^-  &  0 \\ 0 & H^+
\end{array} \right ).
\ee
$H^{\pm}$ are called supersymmetric partner Hamiltonians and  share the
same spectra, apart from the nondegenerate ground state, (see \cite{Cooper1} for a review), 
\be
E_n^{(+)} = E_{n+1}^{(-)}.
\ee
For the non-spontaneously broken supersymmetry this lowest level is of zero energy, $E_1^{(-)} =
0$.  In $\hbar = c = 1$ units, we have
\be
H^{\pm} =  -{1\o 2}{d^2 \o d x^2} + V_{\pm}(x) =  A^{\mp}A^{\pm} 
\ee
where $V_{\pm}(x)$ are called partner potentials. The operators  $A^{\pm}$ are defined in terms
of the superpotential $W(x)$,
\be
\label{As}
A^{\pm} =  {1\o \sqrt 2}\left(\mp {d \o dx} + W(x) \right)
\ee
which satisfies the Riccati equation
\be
\label{Riccati}
W^2 \pm W'=  2V_{\pm}(x)  
\ee
as a consequence of the factorization of the Hamiltonians $H^{\pm}$.

By definition, two partner potentials are called shape invariant if they have the same
functional form, differing only by change of parameters, including an additive constant.
In this case the partner potentials satisfy
\be
V_+(x, a_1) = V_-(x, a_2) + R(a_2),
\ee
where $a_1$ and $a_2$ denote a set of parameters, with $a_2$ being a function of $a_1$,
\be
a_2 = f(a_1)
\ee  
and $R(a_2)$ is independent of $x$.

Through the super-algebra, for a given  Hamiltonian $H_1$, factorized  in terms of the 
bosonic operators,  it is possible  to  construct its hierarchy of  Hamiltonians.   For the
general spontaneously broken supersymmetric case we have
\be
H_1 =  -{1\o 2}{d^2 \o d x^2} + V_1(x) =  A_1^+A_1^-  + E_0^{(1)}
\ee
where $ E_0^{(1)}$ is the lowest eigenvalue.

The bosonic operators are  defined by (\ref{As}) whereas the superpotential $W_1(r)$ satisfies
the Riccati equation
\be
\label{Riccati}
W_1^2 - W_1'=  2V_1(x) - 2E_0^{(1)}.
\ee

The eigenfunction for the lowest state is related to the superpotential
$W_1$ by
\be
\label{eigenfunction}
\Psi_0^{(1)} (x) = N exp( -\int_0^x W_1(\bar x) d\bar x).
\ee
The supersymmetric partner Hamiltonian is given by
\be
\label{partner}
H_2 = A_1^-A_1^+ + E_0^{(1)} =  -{1\o 2}{d^2 \o d r^2} + {1\o 2}(W_1^2 +
W_1')+ E_0^{(1)} .
\ee
Thus, factorizing  $H_2$ in terms of a new pair of bosonic operators,
$A_2^{\pm}$ we get,
\be
\label{H2}
H_2 = A_2^+A_2^- + E_0^{(2)} =  -{1\o 2}{d^2 \o d x^2} + {1\o 2}(W_2^2 -
W_2')+ E_0^{(2)}
\ee
where $E_0^{(2)} $ is the lowest eigenvalue of $H_2$ and $W_2$ satisfy the
Riccati equation,
\be
W_2^2 - W_2'=  2V_2(x) - 2E_0^{(2)}  .
\ee
Thus a whole hierarchy of Hamiltonians can be constructed, with simple
relations connecting the eigenvalues and eigenfunctions of the $n$-members,
\cite{Cooper1}, \cite{Sukumar}, 
\be
H_n = A_n^+A_n^- + E_0^{(n)}
\ee
\be
\label{An}
A_n^{\pm} =  {1\o \sqrt 2}\left(\mp {d \o dx} + W_n(x) \right)
\ee
\be
\label{Psin}
\Psi_n^{(1)} = A_1^+A_2^+...A_n^+\psi_0^{(n+1)}\;,\;\;\;\;E_n^{(1)} = E_0^{(n+1)}
\ee
where $\Psi_0^{(1)} (x) $ is given by (\ref{eigenfunction}).\\

\noindent {\bf  III. The Eckart-type potential}\\

Consider the following  one-dimensional potential, a generalization of the 
Hulth\'en potential,
\be
\label{V1}
V_1(x) = a_1(a_1-\d){ e^{-2\d x}\o 2(1-e^{-\d x})^2} - a_1(2b_1 + \d){e^{-\d x}\o 2(1-e^{-\d
x})} .
\ee

Through the superalgebra the following superpotential 
\be
\label{W1}
W_1 = - a_1 { e^{-\d x}\o 1-e^{-\d x} }+ b_1
\ee
satisfies the  associated Riccati equation (\ref{Riccati}), 
\br
\label{Ricatti}
W_1^2(x) - W_1^{'}(x) & = & 2 V_1 - 2 E_0^{(1)} \\
& = & a_1(a_1 - \d) {e^{-2\d x}\o (1-e^{-\d x})^2} - a_1 (2b_1 + \d)
{e^{-\d x}\o (1-e^{-\d x})} + b_1^2.\nonumber
\er
Thus, substituting (\ref{V1}) into (\ref{Ricatti}) we find that the lowest
energy-eigenvalue is given by
\be
E_0^{(1)} = - {b_1^2 \o 2}
\ee
and from equations (\ref{eigenfunction}) and (\ref{W1}) the associated  eigenfunction is 
given by
\be
\Psi_0^{(1)} = (1 - e^{-\d x})^{{a_1 \o \d}} e^{-b_1 x}.
\ee
The condition we must  impose on  the above wave-function is to
vanish at infinity and at the origin, i.e.,
\be
{a_1 \o \d} > 0\;,\;\;\;\;\; b_1 > 0 
\ee
and since $ \d > 0$ it implies that
\be
a_1 > 0\;,\;\;\;\;\; b_1  > 0.
\ee 
Thus we can evaluate the whole hierarchy of this new potential. 
The supersymmetric partner of
$V_1(x)$ comes from equation (\ref{partner})
\be
W_1^2(x) + W_1^{'}(x) = 2 V_2 - 2 E_0^{(1)}
\ee
and is given by
\be
\label{V2}
V_2(x) = a_1(a_1 + \d){ e^{-2\d x}\o  2(1-e^{-\d x})^2} - a_1(2b_1 - \d){e^{-\d x}\o 2(1-e^{-\d
x})} .
\ee
Considering now the factorization of the  Hamiltonian associated to $V_2(x)$, 
(\ref{H2}), it depends on the superpotential $W_2$ which is of the following form
\be
W_2(x) = - a_2 { e^{-\d x}\o 1-e^{-\d x} }+ b_2.
\ee
The Riccati equation it satisfies is given by
\br
W_2^2(x) - W_2^{'}(x) & = & 2 V_2 - 2 E_0^{(2)}\\
& = & a_2(a_2 - \d) {e^{-2\d x}\o (1-e^{-\d x})^2} - a_2 (2b_2 + \d)
{e^{-\d x}\o (1-e^{-\d
x})} + b_2^2.\nonumber
\er
Again the comparison of the two  equation  above with the
substitution of (\ref{V2}) gives
\be
E_0^{(2)} = - {b_2^2 \o 2}
\ee
where the coefficients are such that
\be
a_2 = a_1 + \d \;,\;\:\;\;\;\;\;\;b_2 = {a_1(2b_1 - \d) \o 2(a_1 + \d)} - {\d
\o 2}.
\ee
\\
Written in terms of $a_2$ and $b_2$ the potential (\ref{V2}) is given by
\be
V_2(x) = a_2(a_2 - \d){ e^{-2\d x}\o  2(1-e^{-\d x})^2} - a_2(2b_2 + \d){e^{-\d x}\o 2(1-e^{-\d
x})} .
\ee
which is analogous to the potential $V_1$ given by equation (\ref{V1}).\\

Thus, this process can be repeated sistematically $n$ times in order to derive the whole
hierarchy. We arrive at the $n$-th term, given by
\be
\label{NewV}
V_{n}(x) = a_{n}(a_{n} -\d){ e^{-2\d x}\o 2(1-e^{-\d x})^2} -
a_{n}(2b_{n} + \d){e^{-\d x}\o 2(1-e^{-\d x})}
\ee
\be
\label{hierarchy}
W_n(x)= - a_n { e^{-\d x}\o 1-e^{-\d x} }+ b_{n}   
\ee
\be
E_0^{(n)} = - {b_n^2 \o 2}  
\ee \\
and the ground state wave function of each member of the hierarchy given by
\be
\label{Psin0}
\Psi_0^{(n)} = (1 - e^{-\d x})^{{a_n \o \d}} e^{-b_n x}
\ee
where the constants satisfy
\be
\label{as}
a_n = a_{n-1} + \d  
\ee
and
\be
b_n = {a_{n-1} (2b_{n-1} - \d) \o 2a_n}  - {\d \o 2}.
\ee
After some algebraic manipulations and using induction arguments we arrive at the
following recurrence relations for the a's and b's
\be
\label{bs}
b_{n+1} = {1 \o 2a_{n+1}} \left ( a_1 (2b_1 + \d) - 2 \d (na_1 + {n(n-1)\d
\o 2}) \right )  - {\d \o 2}.
\ee
For future analysis it is convenient to rewrite the above coefficients like
\be
b_{n+1} = {1 \o 2a_{n+1}}  a_1 (a_1 + 2b_1) -  {1 \o 2}a_{n+1}
\ee
where
\be
a_{n+1} = a_1 + n \d.
\ee
\\
Thus, the full hierarchy for this new potential is completely determined in terms of the
parameters $a_1$, $b_1$ and $\d$. \\

At this point we remark that  the new potential is shape invariant, since all the
potentials preserve the shape in the hierarchy eq. (\ref{NewV}).  To prove it we consider the  
potential $V_- = V_1 - E_0^{(1)}$,  given by
\be
\label{V_-}
V_-(a_1, b_1, x) = a_1(a_1-\d){ e^{-2\d x}\o 2(1-e^{-\d x})^2} - a_1(2b_1 + \d){e^{-\d x}\o
2(1-e^{-\d x})} + {b_1^2 \o 2}.
\ee
Its supersymmetric partner is $V_+(a_1, b_1, x) = V_2(a_1, b_1, x) - E_0^{(1)}$, i.e.,
\be
V_+(a_1, b_1, x) = a_1(a_1 + \d){ e^{-2\d x}\o  2(1-e^{-\d x})^2} - a_1(2b_1 - \d){e^{-\d x}\o
2(1-e^{-\d x})}  + {b_1^2 \o 2}.
\ee
Thus, if the parameters are transformed as 
\be
\label{a2}
a_1 \rightarrow a_2 = a_1 + \d\;,\;\;\; b_1 \rightarrow b_2 = {a_1(2b_1 - \d) \o 2 (a_1 + \d)}-
{\d
\o 2}
\ee
we get
\br
V_-(a_2, b_2, x) &=& a_2(a_2-\d){ e^{-2\d x}\o 2(1-e^{-\d x})^2} - a_2(2b_2 + \d){e^{-\d x}\o
2(1-e^{-\d x})} + {b_2^2 \o 2} \nonumber \\
&=& a_1(a_1 + \d){ e^{-2\d x}\o  2(1-e^{-\d x})^2} - a_1(2b_1 -
\d){e^{-\d x}\o 2(1-e^{-\d x})}  + \nonumber \\
&+& {1\o 2}({a_1(2b_1 - \d) \o 2 (a_1 + \d)}- {\d \o 2})^2.
\er
Therefore we have the identification, in the usual notation
\be
V_+(a_1, b_1, x) = V_-(a_2, b_2, x) + R(a_2, b_2, x) 
\ee
where $a_2$ and $b_2$ are defined in (\ref{a2}). This is precisely
the shape invariant condition, \cite{Gedenshtein}.

Thus the spectrum of $V_-(a_1, b_1, x) $ is
\br
\label{energy}
E_{n+1}^{(1)} &=& - {b_{n+1}^2 \o 2} + {b_1^2 \o 2} \nonumber \\
&=&  -{(a_1^2 - a_{n+1}^2 + 2a_1 b_1)^2 \o 8 a_{n+1}^2} + {b_1^2 \o 2} \;, \;\; n= 0, 1, ...\;.
\er
\\
\noindent {\bf  IV. The hyperbolic functions: the Eckart potential}\\

The Eckart potential has the following form, \cite{Cooper1}
\be
\label{Eckart}
V_E(x) = a^2 + {b^2 \o a^2} + a(a-\alpha) cosech^2 \alpha x - 2b \;coth \alpha x.
\ee
As it is well known it is exactly solvable and shape-invariant. The superpotential associated
through the super-algebra is given by
\be
\label{WE}
W_E  = - a \;coth \alpha x + {b \o a}\;,\;\;\; b>a^2\; .
\ee
Substituting the hyperbolic functions by exponentials we arrive at
\be
V_E(x) = (a - {b \o a})^2 + 4a(a-\alpha) {e^{-2\alpha x} \o 2 (1-e^{-2\alpha x})^2}  - 4b 
{e^{-2\alpha x} \o  2 (1-e^{-2\alpha x})}
\ee
and
\be
W_E  = - {b \o a} + a - {2a \o  (1-e^{-2\alpha x})}.
\ee
The energy eigenstates of the Eckart potential are given by
\be
E_{n+1}^{(1)} = {1 \o 2} \left ( a^2 - (a + n \alpha)^2 + {b^2 \o a^2} -  {b^2 \o (a + n
\alpha)^2} \right )\;, \;\;\; n = 0, 1, ...\;.
\ee
Thus, setting $a = {a_1 \o 2}$, $4b  = a_1(a_1 + 2b_1)$ and $\alpha = {\d \o 2}$ we recover our
starting potential (\ref{V_-}), the superpotential (\ref{W1}) and the energy spectrum
(\ref{energy}).\\

\noindent {\bf  V. A particular case: the Hulth\'en potential}\\

The particular case of setting $a_1= \d$ and $2b_1 +\d = 2$ in (\ref{V1}) reduces $V_1(x)$
to the  known Hulth\'en potential
\be
\label{Hulthen}
V_1(x) = V_H = -\d {e^{-\d x}\o 1-e^{-\d x}}.
\ee
In this case the parameters $a_n $ and $b_n$ reduce to 
\be
a_n = n \d\;,\;\;\; b_n = {1 \o n} - { n \d \o 2}.
\ee
Substituting them in the hierarchy of potentials, equation (\ref{NewV}), we arrive at
\be
\label{VnHulthen}
V_n(x) = {n(n-1) {\d}^2 e^{-2\d x}\o 2 (1-e^{-\d x})^2  } -  {\d(2 + n(1-n)
\d)e^{-\d x} \o  2 (1-e^{-\d x})}.
\ee
The lowest energy levels of the spectrum of energy of the potential (\ref{Hulthen}) are given by
\be
E_n^{(1)} = E_0^{(n)} = - {1 \o 2}({1\o n} -  {n \d \o 2})^2\;, \;\;\; n = 1, ...\;.
\ee
These results are exactly the same obtained to the hierarchy of the Hulth\'en potential in 
\cite{Drigo4}. It is not shape invariant as we can see  from the functional form of $V_1$ and
$V_2$ in equation (\ref{VnHulthen}).\\

\noindent {\bf VI. Conclusions}\\

When dealing with the variational method associated with SQM we found a new
potential that depends on tree parameters (a, b, $\delta$). It is a generalisation of the
Hulth\'en potential. Written in terms of hyperbolic functions it is the  known exactly solvable
and shape invariant Eckart potential.

In conclusion, we remark that the results presented here allowed us to get an unified
description of the Hulth\'en and the Eckart potentials. Through the superalgebra the spectrum and
the hierarchy of both associated Hamiltonians were put together in the same formalism based on
SQM.\\
    		       
\noindent {\bf Acknowledgements}\\

The authors would like to thank Prof. S. Salam\'o and Prof. S. Codriansky for useful comments.

\end{document}